%% file: main.tex
\documentclass[ALICE,manyauthors]{cernphprep}

\newcommand{\pT}{\ensuremath{p_{\rm T}}\xspace}

\newcommand{\pTjet}{\ensuremath{p_{\mathrm{T,\;jet}}}}
\newcommand{\pTchjet}{\ensuremath{p_{\mathrm{T}}^{\mathrm{ch\; jet}}}}

\newcommand{\pTdet}{\ensuremath{p_{\mathrm{T,det}}^{\mathrm{ch\; jet}}}}

\newcommand{\Rmax}{\ensuremath{R_\mathrm{max}}}

\newcommand{\ang}{\ensuremath{\lambda_{\alpha}}}

\newcommand{\angsd}{\ensuremath{\lambda_{\alpha,\mathrm{g}}}}

\newcommand{\mjet}{\ensuremath{m_\mathrm{jet}}}
\newcommand{\mjetsd}{\ensuremath{m_\mathrm{jet,g}}}

\usepackage[comma,square,numbers,sort&compress]{natbib}
\usepackage{hyperref}
\usepackage{lineno}
\usepackage{xspace}
\usepackage{CJKutf8}   
\usepackage[T1]{fontenc}
\usepackage{orcidlink}

\begin{document}
\input{commands.tex}

\begin{titlepage}
\PHyear{2024}       
\PHnumber{283}      
\PHdate{28 October}  

\title{Medium-induced modification of groomed and ungroomed jet mass and angularities in \PbPb{} collisions at $\mathbf{\sqrt{\textit{s}_{\textup{NN}}} = 5.02}$ \TeV{}}
\ShortTitle{Groomed and ungroomed jet mass and angularities in \PbPb{} and \pp{}}   

\Collaboration{ALICE Collaboration\thanks{See Appendix~\ref{app:collab} for the list of collaboration members}}
\ShortAuthor{ALICE Collaboration} 

\begin{abstract}
The ALICE Collaboration presents a new suite of jet substructure measurements in \PbPb{} and \pp{} collisions at a center-of-mass energy per nucleon pair $\snn = 5.02$ \TeV{}. These measurements provide access to the internal structure of jets via the momentum and angle of their constituents, probing how the quark--gluon plasma modifies jets, an effect known as jet quenching. Jet grooming additionally removes soft wide-angle radiation to enhance perturbative accuracy and reduce experimental uncertainties. We report the groomed and ungroomed jet mass $\mjet$ and jet angularities $\lambda_\alpha^\kappa$ using $\kappa=1$ and $\alpha>0$.
Charged-particle jets are reconstructed at midrapidity using the anti-$k_\mathrm{T}$ algorithm with resolution parameter $R = 0.2$. A narrowing of the jet mass and angularity distributions in \PbPb{} collisions with respect to \pp{} is observed and is enhanced for groomed results, confirming modification of the jet core. By using consistent jet definitions and kinematic cuts between the mass and angularities for the first time, previous inconsistencies in the interpretation of quenching measurements are resolved, rectifying a hurdle for understanding how jet quenching arises from first principles and highlighting the importance of a well-controlled baseline. These results are compared with a variety of theoretical models of jet quenching, providing constraints on jet energy-loss mechanisms in the quark--gluon plasma.

\end{abstract}
\end{titlepage}

\setcounter{page}{2} 


\input{01_Introduction}

\input{02_Experimental_setup_and_datasets}

\input{03_Analysis_method}

\input{04_Systematic_uncertainties}

\input{05_Results}

\input{06_Conclusions}


\newenvironment{acknowledgement}{\relax}{\relax}
\begin{acknowledgement}
\section*{Acknowledgements}

We gratefully acknowledge
Daniel Pablos Alfonso, Krishna Rajagopal,
Yasuki Tachibana, Abhijit Majumder,
Shi-Yong Chen (\begin{CJK*}{UTF8}{gkai}陈时勇\end{CJK*}),
and Ben-Wei Zhang (\begin{CJK*}{UTF8}{gkai}张本威\end{CJK*})
for providing theoretical predictions for these results
and for engaging in many useful discussions.

\input{fa_2024-09-12_Opt_C.tex}
\end{acknowledgement}

\bibliographystyle{utphys}   
\bibliography{bibliography}

\newpage
\appendix

%
%

\section{The ALICE Collaboration}
\label{app:collab}
\input{Alice_Authorlist_2024-09-12_Opt_C.tex}
\end{document}

%% file: commands.tex
%

\newcommand{\pp}           {pp\xspace}
\newcommand{\ppbar}        {\mbox{$\mathrm {p\overline{p}}$}\xspace}
\newcommand{\XeXe}         {\mbox{Xe--Xe}\xspace}
\newcommand{\PbPb}         {\mbox{Pb--Pb}\xspace}
\newcommand{\pA}           {\mbox{pA}\xspace}
\newcommand{\pPb}          {\mbox{p--Pb}\xspace}
\newcommand{\AuAu}         {\mbox{Au--Au}\xspace}
\newcommand{\dAu}          {\mbox{d--Au}\xspace}

\newcommand{\s}            {\ensuremath{\sqrt{s}}\xspace}
\newcommand{\snn}          {\ensuremath{\sqrt{s_{\mathrm{NN}}}}\xspace}
\newcommand{\pt}           {\ensuremath{p_{\rm T}}\xspace}
\newcommand{\meanpt}       {$\langle p_{\mathrm{T}}\rangle$\xspace}
\newcommand{\ycms}         {\ensuremath{y_{\rm CMS}}\xspace}
\newcommand{\ylab}         {\ensuremath{y_{\rm lab}}\xspace}
\newcommand{\etarange}[1]  {\mbox{$\left | \eta \right |~<~#1$}}
\newcommand{\yrange}[1]    {\mbox{$\left | y \right |~<~#1$}}
\newcommand{\dndy}         {\ensuremath{\mathrm{d}N_\mathrm{ch}/\mathrm{d}y}\xspace}
\newcommand{\dndeta}       {\ensuremath{\mathrm{d}N_\mathrm{ch}/\mathrm{d}\eta}\xspace}
\newcommand{\avdndeta}     {\ensuremath{\langle\dndeta\rangle}\xspace}
\newcommand{\dNdy}         {\ensuremath{\mathrm{d}N_\mathrm{ch}/\mathrm{d}y}\xspace}
\newcommand{\Npart}        {\ensuremath{N_\mathrm{part}}\xspace}
\newcommand{\Ncoll}        {\ensuremath{N_\mathrm{coll}}\xspace}
\newcommand{\dEdx}         {\ensuremath{\textrm{d}E/\textrm{d}x}\xspace}
\newcommand{\RpPb}         {\ensuremath{R_{\rm pPb}}\xspace}

\newcommand{\nineH}        {$\sqrt{s}~=~0.9$~Te\kern-.1emV\xspace}
\newcommand{\seven}        {$\sqrt{s}~=~7$~Te\kern-.1emV\xspace}
\newcommand{\twoH}         {$\sqrt{s}~=~0.2$~Te\kern-.1emV\xspace}
\newcommand{\twosevensix}  {$\sqrt{s}~=~2.76$~Te\kern-.1emV\xspace}
\newcommand{\five}         {$\sqrt{s}~=~5.02$~Te\kern-.1emV\xspace}
\newcommand{\twosevensixnn}{$\sqrt{s_{\mathrm{NN}}}~=~2.76$~Te\kern-.1emV\xspace}
\newcommand{\fivenn}       {$\sqrt{s_{\mathrm{NN}}}~=~5.02$~Te\kern-.1emV\xspace}
\newcommand{\LT}           {L{\'e}vy-Tsallis\xspace}
\newcommand{\GeVc}         {Ge\kern-.1emV/$c$\xspace}
\newcommand{\MeVc}         {Me\kern-.1emV/$c$\xspace}
\newcommand{\TeV}          {Te\kern-.1emV\xspace}
\newcommand{\GeV}          {Ge\kern-.1emV\xspace}
\newcommand{\MeV}          {Me\kern-.1emV\xspace}
\newcommand{\GeVmass}      {Ge\kern-.2emV/$c^2$\xspace}
\newcommand{\MeVmass}      {Me\kern-.2emV/$c^2$\xspace}
\newcommand{\lumi}         {\ensuremath{\mathcal{L}}\xspace}

\newcommand{\ITS}          {\rm{ITS}\xspace}
\newcommand{\TOF}          {\rm{TOF}\xspace}
\newcommand{\ZDC}          {\rm{ZDC}\xspace}
\newcommand{\ZDCs}         {\rm{ZDCs}\xspace}
\newcommand{\ZNA}          {\rm{ZNA}\xspace}
\newcommand{\ZNC}          {\rm{ZNC}\xspace}
\newcommand{\SPD}          {\rm{SPD}\xspace}
\newcommand{\SDD}          {\rm{SDD}\xspace}
\newcommand{\SSD}          {\rm{SSD}\xspace}
\newcommand{\TPC}          {\rm{TPC}\xspace}
\newcommand{\TRD}          {\rm{TRD}\xspace}
\newcommand{\VZERO}        {\rm{V0}\xspace}
\newcommand{\VZEROA}       {\rm{V0A}\xspace}
\newcommand{\VZEROC}       {\rm{V0C}\xspace}
\newcommand{\Vdecay} 	   {\ensuremath{V^{0}}\xspace}

\newcommand{\ee}           {\ensuremath{e^{+}e^{-}}} 
\newcommand{\pip}          {\ensuremath{\pi^{+}}\xspace}
\newcommand{\pim}          {\ensuremath{\pi^{-}}\xspace}
\newcommand{\kap}          {\ensuremath{\rm{K}^{+}}\xspace}
\newcommand{\kam}          {\ensuremath{\rm{K}^{-}}\xspace}
\newcommand{\pbar}         {\ensuremath{\rm\overline{p}}\xspace}
\newcommand{\kzero}        {\ensuremath{{\rm K}^{0}_{\rm{S}}}\xspace}
\newcommand{\lmb}          {\ensuremath{\Lambda}\xspace}
\newcommand{\almb}         {\ensuremath{\overline{\Lambda}}\xspace}
\newcommand{\Om}           {\ensuremath{\Omega^-}\xspace}
\newcommand{\Mo}           {\ensuremath{\overline{\Omega}^+}\xspace}
\newcommand{\X}            {\ensuremath{\Xi^-}\xspace}
\newcommand{\Ix}           {\ensuremath{\overline{\Xi}^+}\xspace}
\newcommand{\Xis}          {\ensuremath{\Xi^{\pm}}\xspace}
\newcommand{\Oms}          {\ensuremath{\Omega^{\pm}}\xspace}
\newcommand{\degree}       {\ensuremath{^{\rm o}}\xspace}

%% file: 01_Introduction.tex
\section{Introduction} 

Collisions of ultra-relativistic heavy ions at the Large Hadron Collider (LHC) allow the study of bulk properties in quantum chromodynamics (QCD) at high temperature and density. These collisions produce a strongly-interacting state of matter called the quark--gluon plasma (QGP)~\cite{Bjorken_QGP, TheBigPicture} where quarks and gluons are deconfined from nucleons. The hard scattering of two partons from these collisions forms collimated sprays of particles called jets. As they traverse the QGP, the partonic jets lose energy to the medium and their internal structure is modified, an effect known as jet quenching~\cite{Bjorken_quenching, Appel_quenching, 1990_quenching, 1995_quenching, Cao_2021}. Consequently, jets can probe the structure and evolution of the QGP, and provide information about QGP transport properties, degrees of freedom, and the mechanisms for energy loss, as a function of momentum scale.

Jet substructure observables, which characterize the angular and transverse momentum distributions of the particles which constitute jets, can quantify these QGP quenching effects~\cite{Yacine_2017, Chien:2018dfn}.
For example, the jet invariant mass, $\mjet \equiv \sqrt{E_\mathrm{jet}^2 - p_\mathrm{jet}^2}$, where $E_\mathrm{jet}$ is the jet energy and $p_\mathrm{jet}$ its total momentum, has seen extensive experimental~\cite{CMS_mass_dijet, ALICE_mjet_2017, CMS_mass_ttbar, ATLAS_gr_mass_1, CMS_gr_mass, CMS_mass_diff, ATLAS_mass_bb, ATLAS_gr_mass_2, ATLAS_mass_Z, CMS_mass_t_1, CMS_mass_t_2} and theoretical~\cite{Dasgupta_2013, Marzani_mjet_1, Marzani_mjet_2, NNLO_mjet_gr} study in recent years.
The generalized jet angularities~\cite{Berger_2003_1, Berger_2003_2, Berger_2004, Almeida:2008yp, Larkoski_2014} are another class of such observables, defined as
\begin{equation} \label{ang_eqn}
\lambda_\alpha^\kappa \equiv \sum\limits_{i \in \text{jet}}
\bigg( \frac{p_{\text{T},i}}{\pTjet} \bigg)^\kappa
\bigg( \frac{\Delta R_i}{R} \bigg)^\alpha,
\end{equation}
where $i$ runs over constituents in the jet, $p_\mathrm{T}$ designates transverse momentum, $R$ is the jet resolution parameter, and 
$\Delta R_i
\equiv
\sqrt{(y_\text{jet} - y_i)^2 + (\varphi_\text{jet} - \varphi_i)^2}$
gives the distance between the jet axis and its $i$th constituent in the rapidity ($y$) -- azimuthal angle ($\varphi$) plane.
The continuous parameters $\alpha$ and $\kappa$ define the specific observable, where the $\kappa=1$ and $\alpha>0$ configurations are infrared and collinear (IRC) safe~\cite{Berger_2003}.

Both \mjet{} and $\lambda_\alpha^\kappa$ characterize the jet radial energy profile, with a direct theoretical relation between them,
\begin{equation}\label{eq:thrust-mass}
\lambda_{2}^{1} =
\left( \frac{\mjet}{\pTjet R} \right)^2 + \mathcal{O}[(\lambda_2^1)^2],
\end{equation}
where $\lambda^1_2$ is also called the jet thrust~\cite{thrust_d(r)}, and the last term contains higher-order corrections in \mjet{}~\cite{Kang_2018}. The jet thrust is also related to the jet girth~\cite{girth_def}, $g = \lambda_1^1 R$, with a smaller angular weighting $\alpha$. The ALICE collaboration measured $g$ and \mjet{} in \PbPb{} collisions during LHC Run 1 at nucleon--nucleon center-of-mass energy $\snn = 2.76$ \TeV{}, and compared the results to Monte Carlo models of \pp{} collisions~\cite{ALICE_mjet_2017, ALICE_ang_2018}. Significant quenching modification was observed for $g$, while no significant modification was seen for \mjet{}. Since $g$ and $\mjet$ are theoretically related, this discrepancy was unexpected. These measurements differed in their ranges of \pTjet{}, associated with quenching strength and nonperturbative dependence, as well as the angular weighting $\alpha$, associated with momentum broadening, which both could account for the discrepancy.

This letter presents angularities for the 10\% most central \PbPb{} collisions at $\snn = 5.02$ \TeV{}. A recent measurement of IRC-safe angularities in \pp{} collisions at identical center-of-mass energy is used as a no-quenching baseline~\cite{alice_jet_angularities}. We preserve the notation $\lambda_\alpha \equiv \lambda_\alpha^1$ from this measurement, and compare these angularities with new measurements of \mjet{} using the same \pp{} and \PbPb{} collision data, using equivalent $R$ for the first time to address the girth--mass inconsistency.
The results are reported for background-subtracted charged-particle jets with transverse momenta of $40 < \pTchjet < 60$ \GeVc{}.

Soft drop grooming~\cite{Larkoski:2014wba} is employed to remove soft wide-angle radiation from jets, minimizing the nonperturbative dependence of \mjet{} and \ang{}.
Systematically varying $\pTjet$, $\alpha$, $R$, and grooming for each observable provides coherent
constraints on models of jet quenching.

%% file: 02_Experimental_setup_and_datasets.tex
\section{Experimental setup and datasets} 

A description of the ALICE detector and its performance is given in
Refs.~\cite{ALICE_experiment, ALICE_performance}.
The \pp{} data were collected in 2017 at
$\s = 5.02$ \TeV{} during Run 2 of the CERN LHC~\cite{LHCmachine}
using a minimum-bias trigger, which requires a coincidence of hits
in two forward V0 scintillator detectors~\cite{ALICE_V0}.
The \PbPb{} data were collected during 2018 at $\snn = 5.02$ \TeV{}, using
high-multiplicity events in the V0 detectors to trigger on
the 10\% most central collisions (0--10\% centrality class)~\cite{ALICE_centrality_HI}.
The event selection includes a primary-vertex selection and the
removal of beam-induced background events and pileup~\cite{ALICE_pileup}.
The \pp{} data sample contains 870 million events and corresponds to an
integrated luminosity of $18.0 \pm 0.4$ nb$^{-1}$~\cite{ALICE_luminosity_pp}.
The 0-10\% centrality \PbPb{} data sample contains 91.2 million events,
corresponding to an integrated luminosity of $0.119 \pm 0.003$ nb$^{-1}$~\cite{ALICE_luminosity_Pb-Pb}.

Charged-particle tracks are reconstructed using information from both the Time
Projection Chamber (TPC)~\cite{ALICE_TPC} and the Inner Tracking
System (ITS)~\cite{ALICE_ITS}, using both global tracks and complementary tracks.
Global tracks are required to include at least one hit in the silicon
pixel detector (SPD), comprising the two innermost layers of the ITS,
and to satisfy a number of quality criteria~\cite{ALICE_pp_jet_cross_section}.
Complementary tracks do not contain any hits in the SPD, but otherwise
satisfy the tracking criteria, and are refit using the
primary vertex of the event to constrain their trajectory. Including complementary tracks
ensures approximately uniform azimuthal acceptance, while preserving
similar \pT{} resolution to tracks with SPD hits.
Tracks with $0.15 < \pT < 100$ \GeVc{} are accepted over the pseudorapidity
range $|\eta| < 0.9$. The accepted tracks exhibit a momentum resolution 
ranging from about 1\% at track $\pT = 1$ \GeVc{} to 4\% at
track $\pT = 50$ \GeVc{}.

%% file: 03_Analysis_method.tex
\section{Analysis method} 

Jets are reconstructed from charged-particle tracks with FastJet 3.3.3~\cite{fastjet} using the anti-$k_\mathrm{T}$ algorithm~\cite{antikt} with $E$-scheme recombination and resolution parameter $R = 0.2$.
This small value of $R$ reduces contamination from combinatorial jets at low \pTchjet{} in \PbPb{} data, thereby increasing overlap with the measured vacuum baseline~\cite{alice_jet_angularities}.
Despite track-based observables being collinear-unsafe~\cite{Chang_2013}, they offer greater momentum and angular precision than calorimeter-based observables.
The $\pi^\pm$-meson mass is assumed for all jet constituents, and the jet pseudorapidity $\eta_\mathrm{jet}$ is required to be within the fiducial volume of the TPC, $|\eta_\mathrm{jet}| < 0.9 - R$. For \pp{} collisions, all reconstructed jets in the range $5 < \pTchjet{} < 200$ \GeVc{} are analyzed. In heavy-ion collisions, jets are influenced by a large background from the underlying event (UE)~\cite{ALICE_jet_UE}, owing to the large number of soft, thermally-produced particles from the QGP. To reduce this thermal background, the event-by-event constituent subtraction method is used~\cite{background_sub}, which adds ``ghosts'' to the event over the entire acceptance. These ghosts, whose small transverse momentum is calculated from the average background density, are then combined with real local particles within a maximum recombination distance \Rmax{} = 0.1. When combined, the softer (lower \pt{}) particle of the pair is removed from the event, while its \pt{} and mass is subtracted from that of the harder particle. After background subtraction, the measured range is truncated to $40 < \pTchjet < 200$ \GeVc{} before applying corrections.

Jets are groomed using soft drop~\cite{Larkoski:2014wba}, which reclusters the jet into an angularly-ordered tree using the Cambridge/Aachen algorithm~\cite{cambridge_aachen} before iterating along the hardest branch and trimming away soft, wide-angle radiation at each splitting until the soft drop condition is satisfied, $\pt{}_\mathrm{,min} / (\pt{}_\mathrm{,min} + \pt{}_\mathrm{,max}) > z_\mathrm{cut} (\Delta R / R)^\beta$, where $\pt{}_\mathrm{,i}$ are the transverse momenta of the branches and $z_\mathrm{cut}$ and $\beta$ are user parameters, which are set to 0.2 and 0, respectively. These settings require the jet to have a splitting where the softer branch carries 20\% or more of the total transverse momentum of the splitting (i.e., $z_\mathrm{cut}=0.2$) independent of the angle of the splitting (i.e., $\beta=0$), which improves the efficiency of tagging the first hard splitting in the large background of \PbPb{} collisions~\cite{Mulligan_2020}.

The reconstructed jet mass and jet angularity distributions are affected by tracking inefficiency, particle interactions with detector material, and finite track \pt{} resolution. Moreover, the background subtraction procedure in \PbPb{} collisions only corrects for the average UE, and remaining background fluctuations smear the reconstructed distributions. To account for these effects, \pp{} events are simulated with the PYTHIA 8 generator using the Monash 2013 tune~\cite{PYTHIA_intro} and passed through a setup of the ALICE detector using GEANT3~\cite{GEANT3} to account for the particle transport through the detector material. 
For the \PbPb{} analysis, the PYTHIA 8 simulations including GEANT3 reconstruction (detector level) are embedded into reconstructed \PbPb{} data events (combined level). Background subtraction and jet reconstruction are performed on the combined-level events, identical to the data analysis. Jets are matched geometrically between the detector (in \pp{}) or combined (in \PbPb{}) level to the jets reconstructed from the PYTHIA 8 simulation without detector effects (truth-level), with these jet matches required to be unique. In \PbPb{}, the matched combined-level jet must contain tracks from the detector-level jet amounting to at least 50\% of the \pT{} of the detector-level jet. These requirements allow for a reliable
estimation of background effects and fluctuations on the observables.

A four-dimensional response matrix (RM) describing the detector and background response in \pTchjet{} and \ang{} or \mjet{} is constructed from the jets matched between detector-level (combined-level) and generator-level and used in a two-dimensional unfolding with the iterative Bayesian unfolding algorithm~\cite{Bayes_unfolding} as implemented in RooUnfold~\cite{RooUnfold}. The number of iterations through the unfolding was optimized by ensuring good closure of the unfolding procedure at the earliest iteration, with values ranging from 5 to 15 depending on the collision system and \pTchjet{} interval. More details on the background subtraction and jet matching can be found in Ref.~\cite{ALICE_public_note}.

%% file: 04_Systematic_uncertainties.tex
\section{Systematic uncertainties} 

The systematic uncertainties for the observables reported in this paper are estimated from the uncertainty on the tracking efficiency, the unfolding, and the dependence on the event generator used in the simulations. In \PbPb{}, additional uncertainties arise from the estimation of the thermal background and the background subtraction procedure. Variations to the analysis procedure (described below) are performed to estimate these uncertainties, with the relative variation between unfolded distributions obtained with the default and modified procedures taken as the relative systematic uncertainty. The total systematic uncertainty is then taken as the quadratic sum of all contributions. The procedure is the same with and without grooming.

The systematic uncertainty due to the tracking efficiency uncertainty is evaluated using random rejection of tracks before jet finding. The tracking efficiency uncertainty is estimated to be 3\% in \pp{} and 3--5\% in \PbPb{} collisions, depending on track \pT{}, based on variations in the track selection criteria and on the ITS--TPC track-matching efficiency uncertainty. The RM is recreated after randomly rejecting the fraction of tracks equal to the corresponding uncertainty, and the results are unfolded.
The systematic uncertainty arising from the unfolding regularization is evaluated by varying the number of unfolding iterations by $\pm 2$ units, altering the shape of the prior distribution, changing the detector-level observable binning, and truncating the lower and upper bounds in the detector-level charged-particle jet transverse momentum \pTdet{} range by 5 and 80 \GeVc{}, respectively.
The systematic uncertainty due to the model dependence of the generator used to construct the response matrix is estimated by comparing results obtained via unfolding with RMs generated using PYTHIA 8~\cite{PYTHIA_intro} and HERWIG 7~\cite{herwig_manual, herwig_release} for the \pp{} and \PbPb{} analyses, and also JEWEL~\cite{JEWEL_manual} (recoils off) for the \PbPb{} analysis.
The systematic uncertainty introduced by the background subtraction in Pb--Pb collisions using the constituent subtraction procedure is estimated by varying \Rmax{} from ``under-subtraction'' (\Rmax{} = 0.05) to ``over-subtraction'' (\Rmax{} = 0.5), around the nominal value of \Rmax{} = 0.1. 
The uncertainty due to residual contamination from the uncorrelated thermal background produced by the QGP is estimated by embedding combined-level MC jets into a simulated thermal background and applying the analysis procedure, with any non-closure observed in the unfolding taken as a systematic uncertainty.
The total relative systematic uncertainty ranges from 2\% to 26\% for \ang{} and 3\% to 30\% for \mjet{}, with larger values in the tails of the \ang{} and \mjet{} distributions and lower \pTchjet{} intervals.
See Ref.~\cite{ALICE_public_note} for more details about the systematic uncertainties used in this measurement.

%% file: 05_Results.tex
\section{Results} 

In this letter we report the \mjet{} and \ang{} distributions in the charged-jet transverse momentum interval $40 < \pTchjet{} < 60$ \GeVc{} in inclusive \pp{} and central (0–10\%) \PbPb{} collisions; additional \pTchjet{} ranges from 60 to 150 \GeVc{} are reported in Ref.~\cite{ALICE_public_note}. The inclusive jet angularities from Ref.~\cite{alice_jet_angularities}, measured in pp collisions at $\s=5.02$ \TeV{}, are used as a baseline for these \PbPb{} results. The distributions are reported both with and without soft drop grooming (using the parameters $z_\mathrm{cut} = 0.2$ and $\beta=0$) as normalized differential cross sections,
\begin{equation} \label{eq:cross_section}
\frac{1}{\sigma} \frac{\rm{d}\sigma}{\rm{d}\ang}
\equiv
\frac{1}{N_{\text{jets}}} \frac{\mathrm{d}N_\text{jets}} {\rm{d}\ang}
\text{ (ungroomed),} \hspace{1em} \rm{or} \hspace{1em}
\frac{1}{\sigma_{\rm{inc}}} \frac{\rm{d}\sigma}{\rm{d}\angsd}
\equiv
\frac{1}{\it{N}_{\text{jets}}}
\frac{\mathrm{d}\it{N}_\text{gr jets}} {\rm{d}\angsd} \text{ (groomed),}
\end{equation}
where $N_{\mathrm{jets}}$ is the number of inclusive (ungroomed) jets within a given \pTchjet{}
range and $\sigma$ is the corresponding cross section.
For the groomed case, some jets (including all single-particle jets) are removed by the grooming procedure, and therefore
the cross section is explicitly normalized by the number of inclusive (ungroomed) jets; for the ungroomed case, $\sigma=\sigma_\mathrm{inc}$.
The analog of Eq.~\ref{eq:cross_section} also applies for \mjet{}.

\begin{figure}[t]
\center
\includegraphics[width=0.32\textwidth]{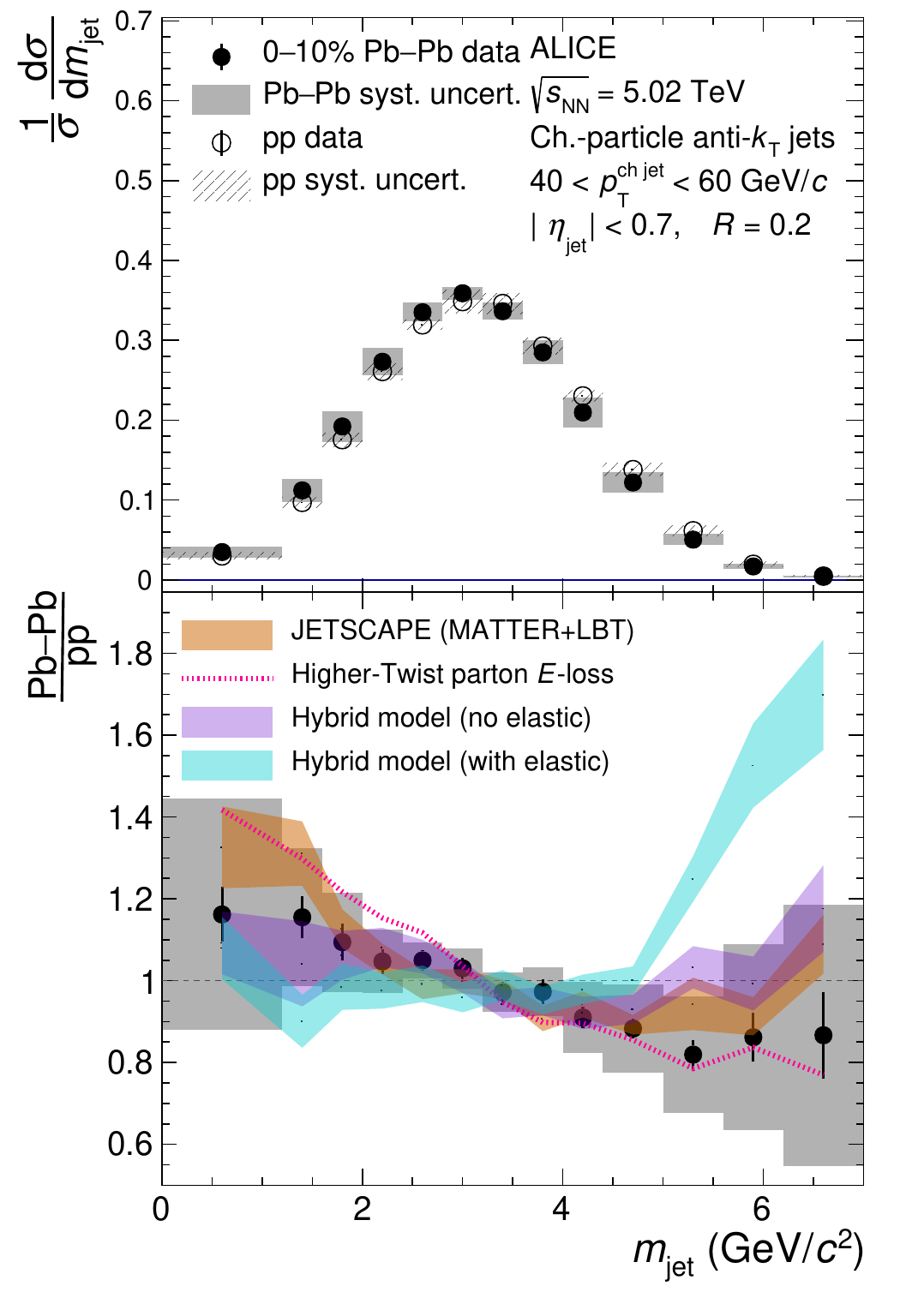}
\includegraphics[width=0.32\textwidth]{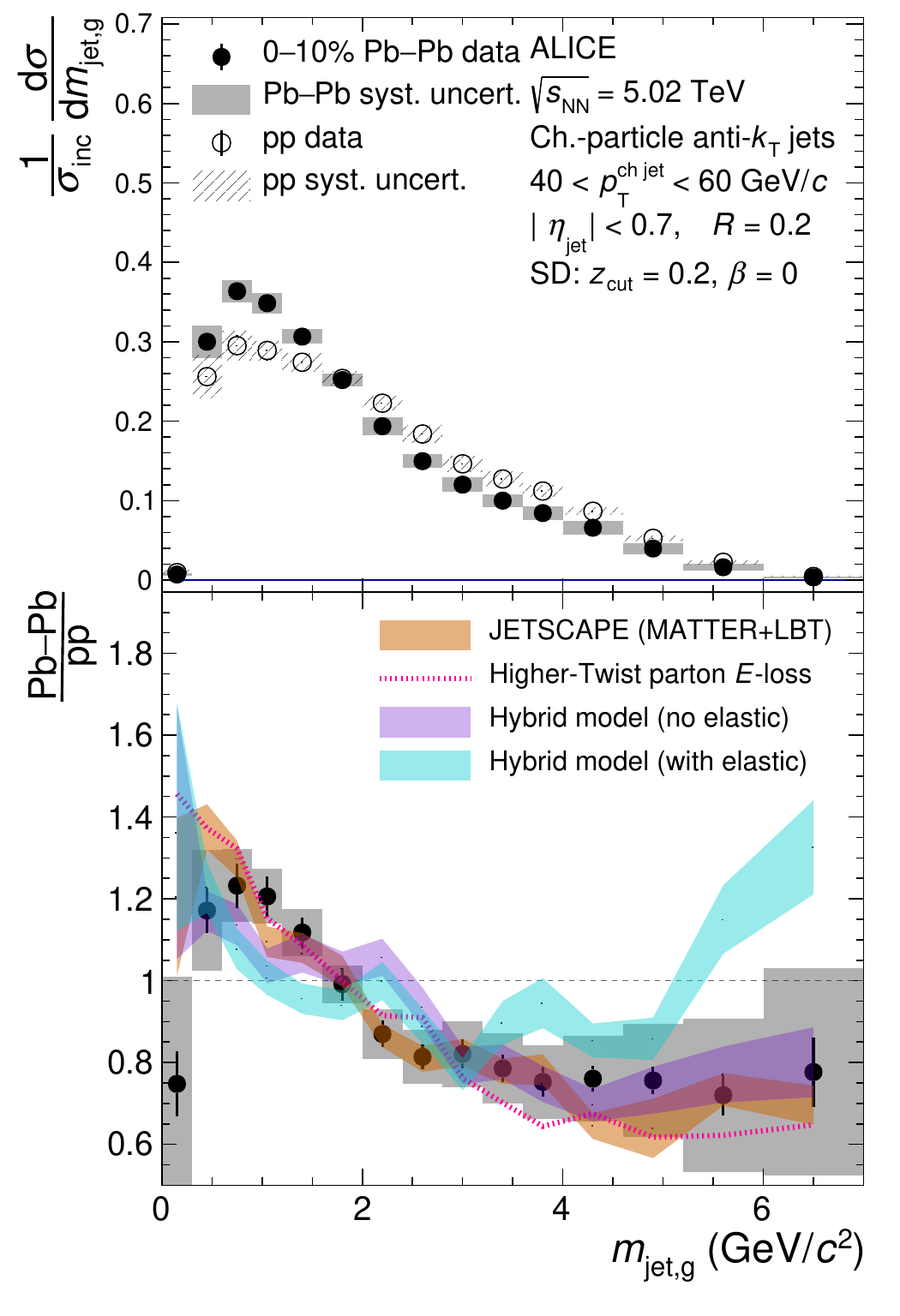}
\caption{ALICE measurements of ungroomed \mjet{} (left) and groomed \mjetsd (right) using $R=0.2$ charged-particle jets with $40 < \pTchjet{} < 60$ \GeVc{} in \pp{} and \PbPb{} collisions at $\snn{} = 5.02$ \TeV{} compared to models. The bottom panel shows the ratio of \PbPb{} distributions to \pp{}, which quantifies the substructure modifications from quenching.}
\label{fig:jet_mass}
\end{figure}

\begin{figure}[t]
\center
\includegraphics[width=0.32\textwidth]{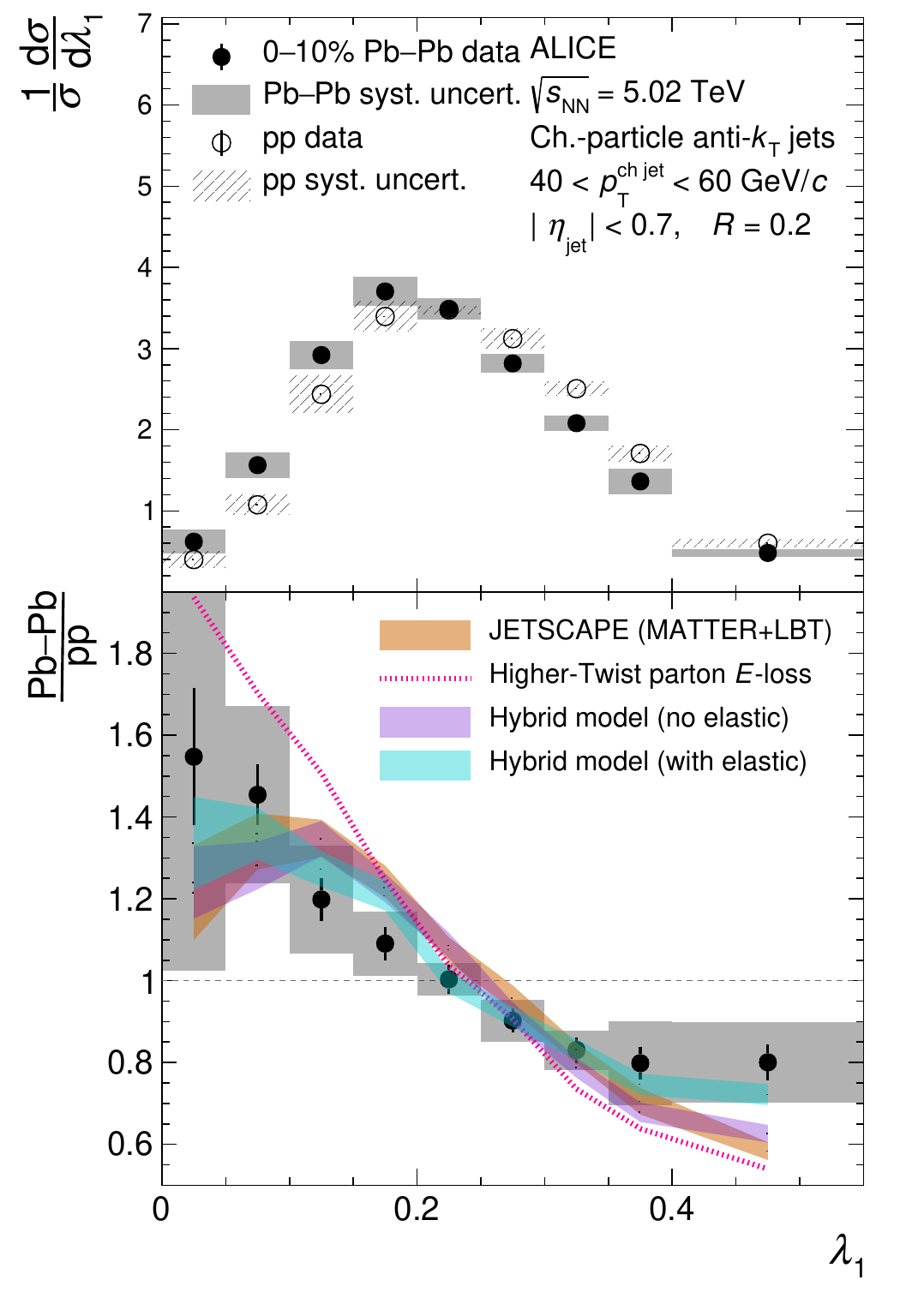}
\includegraphics[width=0.32\textwidth]{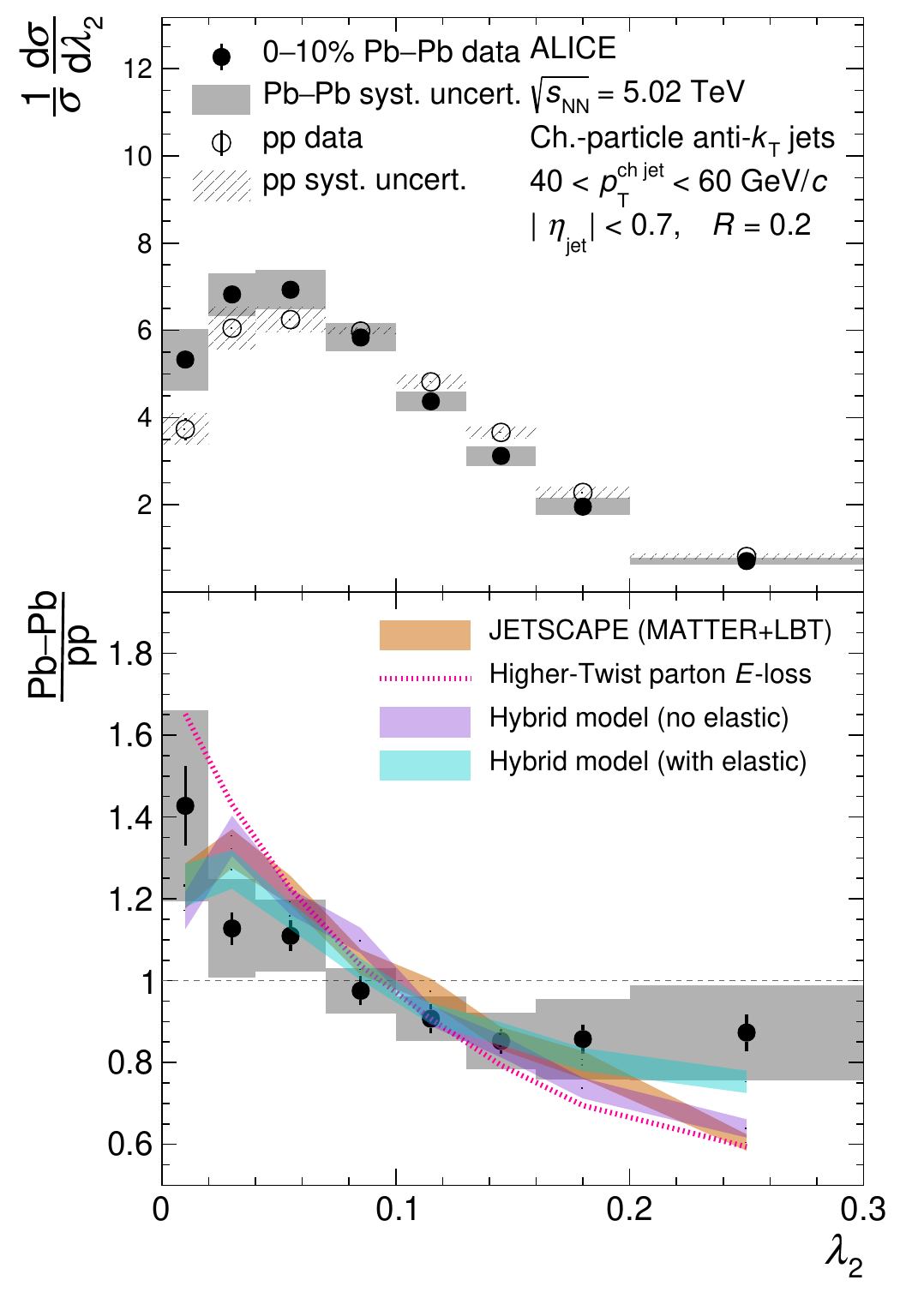}
\includegraphics[width=0.32\textwidth]{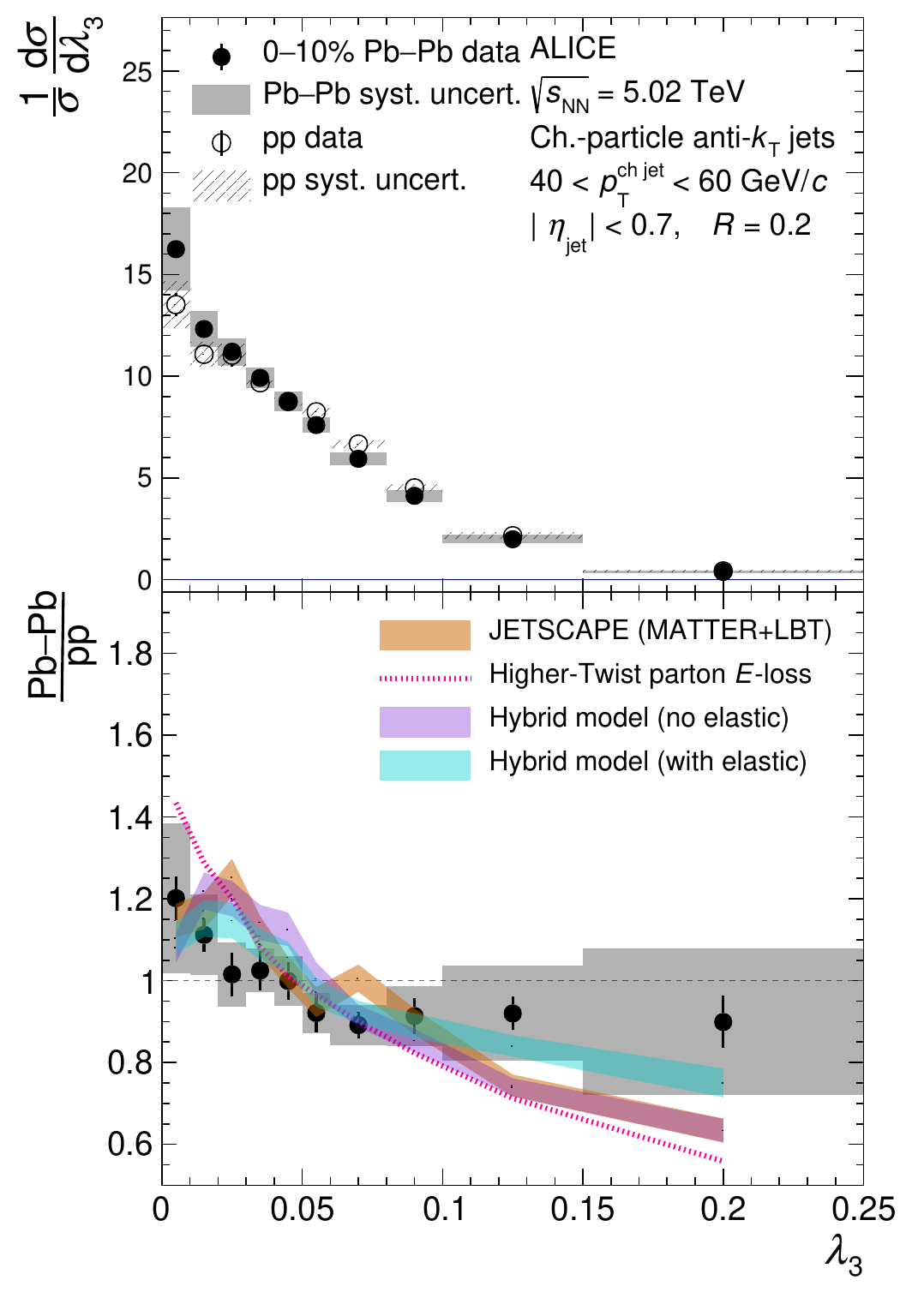}
\caption{ALICE measurements of ungroomed \ang{} for $\alpha=1$ (`girth,' left), $\alpha=2$ (`thrust,' center), and $\alpha=3$ (right) using $R=0.2$ charged-particle jets with $40 < \pTchjet{} < 60$ \GeVc{} in \pp{} and \PbPb{} collisions at $\snn{} = 5.02$ \TeV{} compared to models. The bottom panel shows the ratio of \PbPb{} distributions to \pp{}, which quantifies the substructure modifications from quenching.}
\label{fig:ungroomed_jet_ang}
\end{figure}

\begin{figure}[t]
\center
\includegraphics[width=0.32\textwidth]{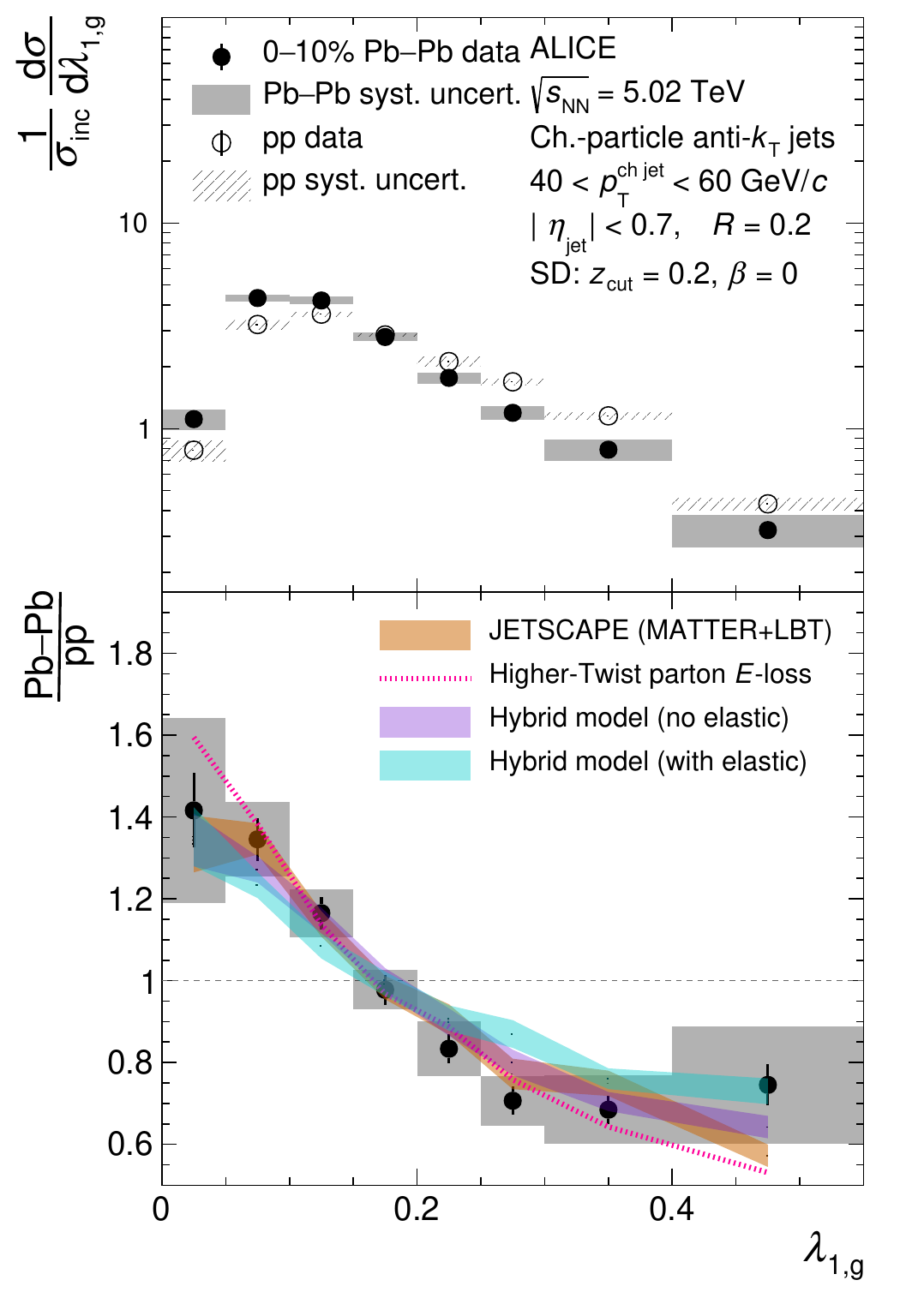}
\includegraphics[width=0.32\textwidth]{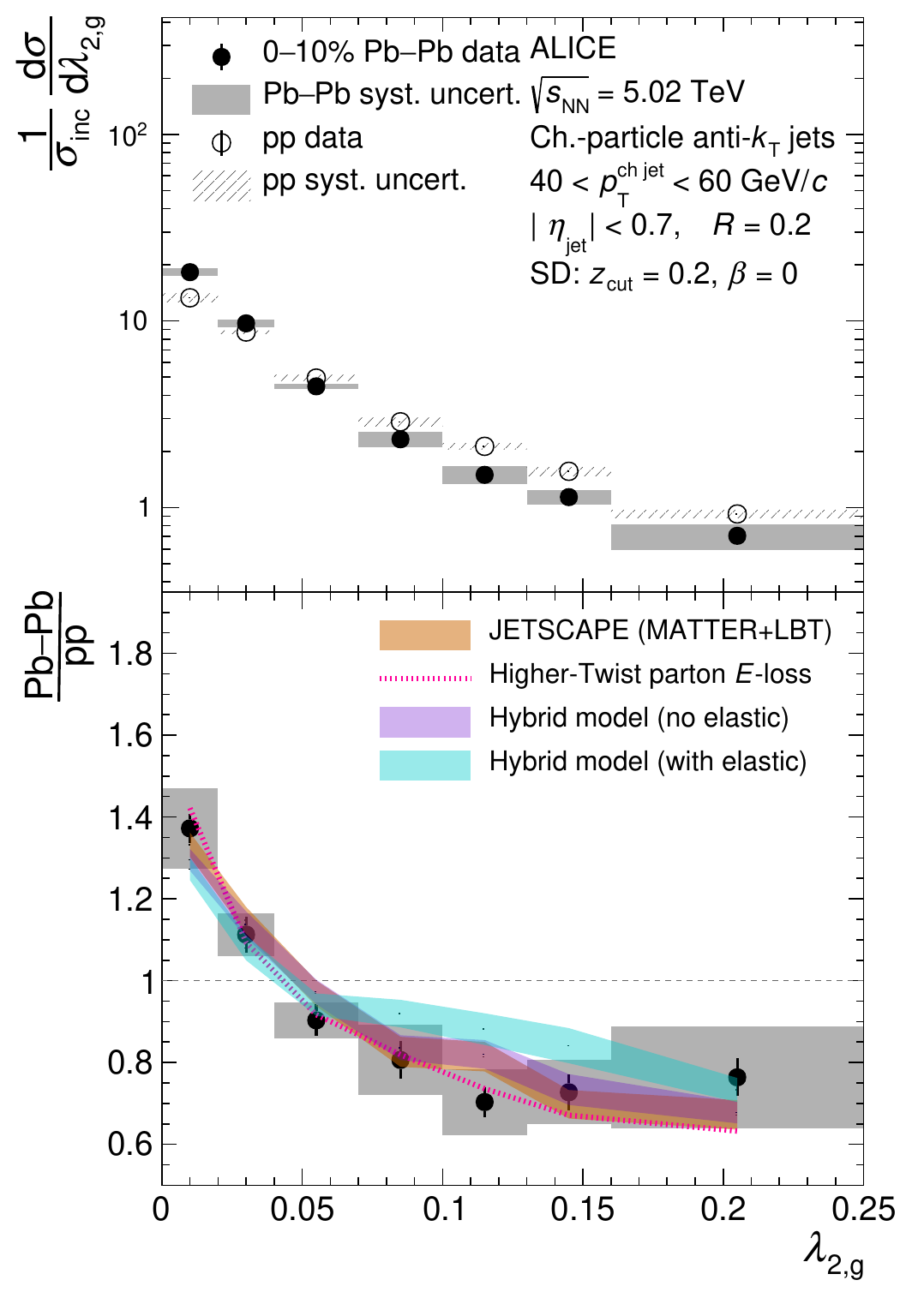}
\includegraphics[width=0.32\textwidth]{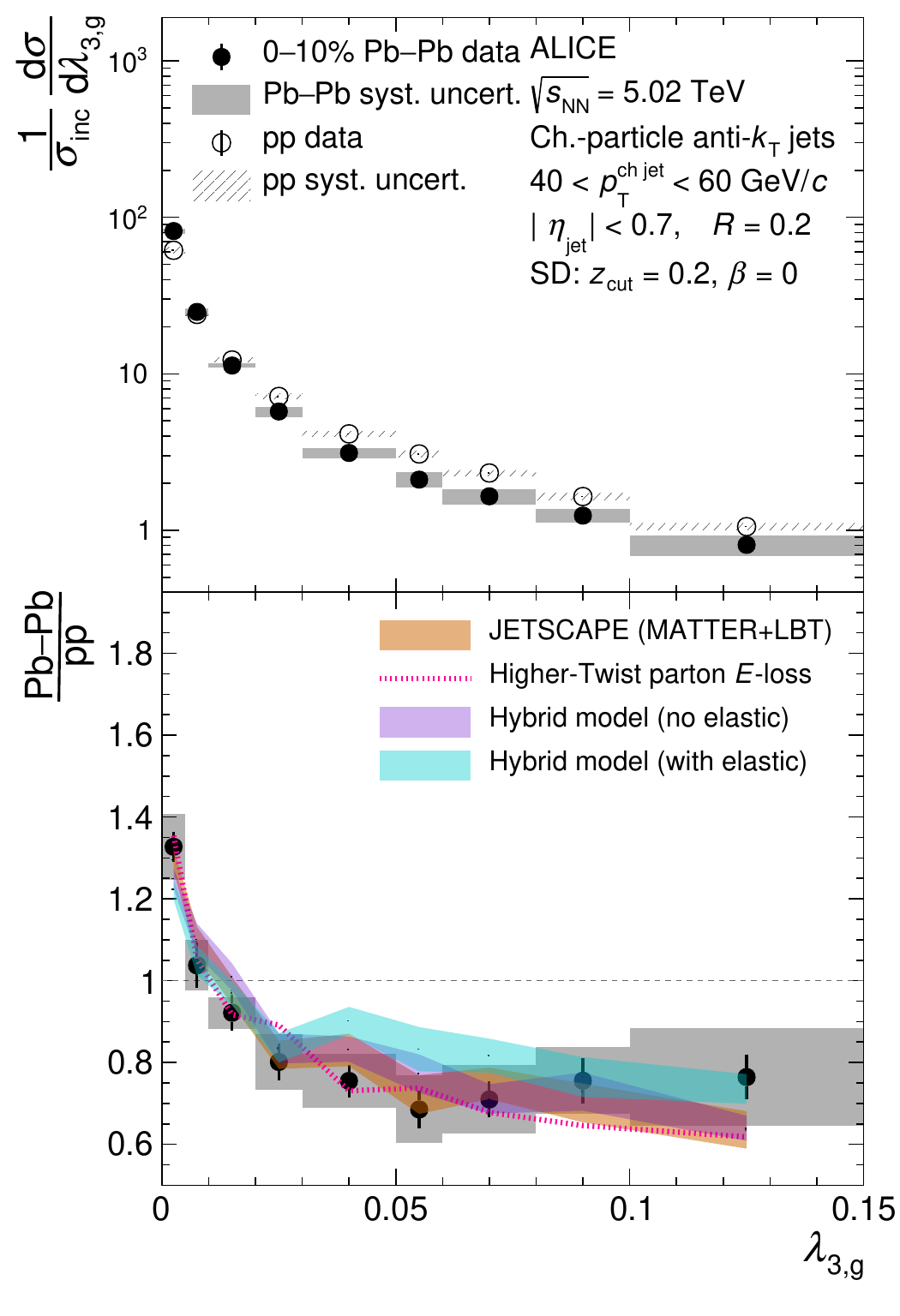}
\caption{ALICE measurements of groomed \angsd{} for $\alpha=1$ (`girth,' left), $\alpha=2$ (`thrust,' center), and $\alpha=3$ (right) using $R=0.2$ charged-particle jets with $40 < \pTchjet{} < 60$ \GeVc{} in \pp{} and \PbPb{} collisions at $\snn{} = 5.02$ \TeV{} compared to models. The bottom panel shows the ratio of Pb–Pb distributions to pp, which quantifies the substructure modifications from quenching. These ratios are visibly enhanced as compared to the ungroomed distributions shown in Fig.~\ref{fig:ungroomed_jet_ang}, signifying a strongly quenched jet core.}
\label{fig:groomed_jet_ang}
\end{figure}

Figure~\ref{fig:jet_mass} shows the ungroomed \mjet{} and groomed \mjetsd{} distributions for $40 < \pTchjet < 60$ \GeVc{}, while the ungroomed \ang{} distributions are shown in Fig.~\ref{fig:ungroomed_jet_ang} and the groomed $\ang$ distributions are displayed in Fig.~\ref{fig:groomed_jet_ang}. The distributions from \PbPb{} collisions are compared to \pp{} data and several models, with $\PbPb/\mathrm{\pp}$ ratios displayed in the bottom panels.
The relative uncertainties are assumed to be uncorrelated between collision systems, and theoretical error bands are purely statistical. 
The ratios suggest an enhancement at small values of angularity and mass and a corresponding suppression at large values, consistent with jet ``narrowing,'' i.e. the transverse momentum becoming more collimated along the jet axis; 
systematic uncertainties, however, are significant, especially for ungroomed \mjet{}.

Quenching modification of the ungroomed jet girth ($\lambda_1$), as quantified in the $\PbPb/\mathrm{\pp}$ ratio, is smaller than when using a MC simulated \pp{} baseline generated with PYTHIA 8, where both tails of the ratio are modified by an approximate factor of 2~\cite{ALICE_ang_2018}. This difference is explained by an approximate 20\% difference between the central values of the \pp{} data and PYTHIA 8 simulation in the tails of the distribution~\cite{ALICE_public_note}. This difference, which enhanced the observed quenching effects in Ref.~\cite{ALICE_ang_2018}, highlights the importance of measuring a proper vacuum baseline for jet quenching. Modification of the jet thrust $\lambda_2$ is more significant than \mjet{}, despite the theoretical relation given in Eq.~\ref{eq:thrust-mass}.

Comparisons to various theoretical calculations are shown. JETSCAPE~\cite{JETSCAPE} uses a medium-modified parton shower described by MATTER~\cite{MATTER} for the high-virtuality phase and Linear Boltzmann Transport (LBT)~\cite{LBT} for the low-virtuality phase. Higher-Twist parton energy loss calculations~\cite{Higher-Twist} use POWHEG~\cite{POWHEG} + PYTHIA 6 as a vacuum baseline; as the partons traverse the QGP, they probabilistically emit medium-induced gluon radiation following the $\hat{q}$-dependent Higher-Twist formalism~\cite{Guo_2000, ZHANG2003429, Zhang_2004, PhysRevD.85.014023}. Finally, the Hybrid model~\cite{Hybrid-Model} allows partons produced by a vacuum shower to undergo medium interactions according to a strongly-coupled AdS/CFT-based model. Two variations are given, where partons do or do not interact via in-medium elastic Moli\`{e}re scattering~\cite{Hybrid-Moliere}, which scatters particles at large angles and broadens both \mjet{} and \ang{}. Individual comparisons of groomed and ungroomed \mjet{} and \ang{} distributions between \pp{} and \PbPb{} data and model predictions are reported in Ref.~\cite{ALICE_public_note}.

Models show general agreement with the measured \PbPb{}/\pp{} ratios within uncertainties. However, data disfavors the Hybrid model with in-medium elastic scattering at large \mjet{} ($2.3 \sigma$ difference in the last bin) but slightly favors it at large \ang{} ($\alpha=2$ exhibits a $0.94 \sigma$ difference in the last bin, as compared to $1.8 \sigma$ for no elastic scattering).
While Eq.~\ref{eq:thrust-mass} relates \mjet{} and $\lambda_2$ directly to one
another, model comparisons show differing behavior within this kinematic regime. 
Since the distributions are positive definite and obey square proportionality following Eq.~\ref{eq:thrust-mass}, large corrections to this equation must apply at these values of \pTchjet{}.
These could include nonperturbative effects such as hadronization as well as higher-order correction terms $\mathcal{O}[(\lambda_2)^2]$, which both could be significant at these smaller values of \pTchjet{} where the strong coupling $\alpha_S$ is large.
Therefore, despite the close mathematical relationship, the observables remain different.
Underlying physical differences also exist, as jet mass is sensitive to quark masses, whereas the IRC-safe jet angularities are more sensitive to the angular profile of jet fragmentation.
The behavioral discrepancies between the measured distributions originate from these physical differences of the observables themselves, which clarifies the girth--mass discrepancy observed between earlier measurements. This observation highlights the importance of making broad measurements of quenched jet substructure, as closely-related observables can provide significantly different probes of underlying physical phenomena.

Increasing the value of $\alpha$ in Fig.~\ref{fig:ungroomed_jet_ang} correspondingly increases the weight of wide-angle jet constituents, which are more affected by nonperturbative processes~\cite{alice_jet_angularities, Kang_2018}.
Less narrowing is observed with increased $\alpha$ for the ungroomed \ang{}, revealing a strongly quenched jet core.
This conclusion is supported by a significant enhancement in the narrowing effect of \ang{} for jets groomed with soft drop, comparing Figs.~\ref{fig:ungroomed_jet_ang} and~\ref{fig:groomed_jet_ang}. 
Removing soft, wide angle jet constituents substantially modifies the shapes of the groomed distributions, with the distributions of \mjetsd{} and \angsd{} peaking at lower values than \mjet{} and \ang{}.
Soft drop grooming also reduces systematic uncertainties via limiting the effects from tracking inefficiency and generator dependence. Nevertheless, most models describe the groomed observables better than the ungroomed ones, despite the smaller uncertainties of the groomed results, as grooming reduces the influence of soft, wide-angle radiation thus improving theoretical control.

%% file: 06_Conclusions.tex
\section{Conclusions} 

In this letter, measurements of jet mass and angularities in \pp{} and \PbPb{} collisions at $\snn = 5.02$ \TeV{} are presented. The medium-induced jet modifications in \PbPb{} collisions are explored both with and without grooming. These results depict a consistent picture of narrowing as jets traverse the QGP, which is dominated by a collinearized jet core. By measuring both \mjet{} and jet thrust ($\lambda_{2}$) using the same jets, and by also measuring the appropriate \pp{} baseline, we reexamine the inconsistency between the girth and mass distributions raised by earlier measurements, which showed conflicting quenching behavior in these related observables. 
Fundamental differences are found between these observables at low \pTchjet{} ($40 < \pTchjet{} < 60$ \GeVc{}) also in the analysis presented in this letter. This indicates that the mass--thrust relation (Eq.~\ref{eq:thrust-mass}) must depend on significant higher-order corrections or on nonperturbative physics at low \pTchjet{}, where the strong coupling $\alpha_S$ is large.

The data generally agree with models including in-medium energy loss. The jet mass prefers no in-medium elastic Moli\`{e}re scattering (within the Hybrid model), but the jet angularities slightly prefer if this process is included. Future studies are needed to understand these model discrepancies, which may be alleviated with further tuning. Theory comparisons also reveal that a \pp{} baseline is essential for evaluating quenching behavior of jet substructure and should always be measured to fully profit from heavy-ion runs at the LHC. Compared to previous measurements using a MC simulated \pp{} baseline, the quenching effects are smaller.

While jet grooming has been used in many recent measurements, the phase space of groomed observables remains mostly unexplored. Using grooming to reduce experimental uncertainties while selecting observables which probe effects such as in-medium color coherence will be essential to illuminate medium structure and the origins of jet quenching. Grooming can also be used to reduce nonperturbative effects, providing a handle to isolate these mechanisms in the QGP.

%% file: fa_2024-09-12_Opt_C.tex

The ALICE Collaboration would like to thank all its engineers and technicians for their invaluable contributions to the construction of the experiment and the CERN accelerator teams for the outstanding performance of the LHC complex.
The ALICE Collaboration gratefully acknowledges the resources and support provided by all Grid centres and the Worldwide LHC Computing Grid (WLCG) collaboration.
The ALICE Collaboration acknowledges the following funding agencies for their support in building and running the ALICE detector:
A. I. Alikhanyan National Science Laboratory (Yerevan Physics Institute) Foundation (ANSL), State Committee of Science and World Federation of Scientists (WFS), Armenia;
Austrian Academy of Sciences, Austrian Science Fund (FWF): [M 2467-N36] and Nationalstiftung f\"{u}r Forschung, Technologie und Entwicklung, Austria;
Ministry of Communications and High Technologies, National Nuclear Research Center, Azerbaijan;
Conselho Nacional de Desenvolvimento Cient\'{\i}fico e Tecnol\'{o}gico (CNPq), Financiadora de Estudos e Projetos (Finep), Funda\c{c}\~{a}o de Amparo \`{a} Pesquisa do Estado de S\~{a}o Paulo (FAPESP) and Universidade Federal do Rio Grande do Sul (UFRGS), Brazil;
Bulgarian Ministry of Education and Science, within the National Roadmap for Research Infrastructures 2020-2027 (object CERN), Bulgaria;
Ministry of Education of China (MOEC) , Ministry of Science \& Technology of China (MSTC) and National Natural Science Foundation of China (NSFC), China;
Ministry of Science and Education and Croatian Science Foundation, Croatia;
Centro de Aplicaciones Tecnol\'{o}gicas y Desarrollo Nuclear (CEADEN), Cubaenerg\'{\i}a, Cuba;
Ministry of Education, Youth and Sports of the Czech Republic, Czech Republic;
The Danish Council for Independent Research | Natural Sciences, the VILLUM FONDEN and Danish National Research Foundation (DNRF), Denmark;
Helsinki Institute of Physics (HIP), Finland;
Commissariat \`{a} l'Energie Atomique (CEA) and Institut National de Physique Nucl\'{e}aire et de Physique des Particules (IN2P3) and Centre National de la Recherche Scientifique (CNRS), France;
Bundesministerium f\"{u}r Bildung und Forschung (BMBF) and GSI Helmholtzzentrum f\"{u}r Schwerionenforschung GmbH, Germany;
General Secretariat for Research and Technology, Ministry of Education, Research and Religions, Greece;
National Research, Development and Innovation Office, Hungary;
Department of Atomic Energy Government of India (DAE), Department of Science and Technology, Government of India (DST), University Grants Commission, Government of India (UGC) and Council of Scientific and Industrial Research (CSIR), India;
National Research and Innovation Agency - BRIN, Indonesia;
Istituto Nazionale di Fisica Nucleare (INFN), Italy;
Japanese Ministry of Education, Culture, Sports, Science and Technology (MEXT) and Japan Society for the Promotion of Science (JSPS) KAKENHI, Japan;
Consejo Nacional de Ciencia (CONACYT) y Tecnolog\'{i}a, through Fondo de Cooperaci\'{o}n Internacional en Ciencia y Tecnolog\'{i}a (FONCICYT) and Direcci\'{o}n General de Asuntos del Personal Academico (DGAPA), Mexico;
Nederlandse Organisatie voor Wetenschappelijk Onderzoek (NWO), Netherlands;
The Research Council of Norway, Norway;
Pontificia Universidad Cat\'{o}lica del Per\'{u}, Peru;
Ministry of Science and Higher Education, National Science Centre and WUT ID-UB, Poland;
Korea Institute of Science and Technology Information and National Research Foundation of Korea (NRF), Republic of Korea;
Ministry of Education and Scientific Research, Institute of Atomic Physics, Ministry of Research and Innovation and Institute of Atomic Physics and Universitatea Nationala de Stiinta si Tehnologie Politehnica Bucuresti, Romania;
Ministry of Education, Science, Research and Sport of the Slovak Republic, Slovakia;
National Research Foundation of South Africa, South Africa;
Swedish Research Council (VR) and Knut \& Alice Wallenberg Foundation (KAW), Sweden;
European Organization for Nuclear Research, Switzerland;
Suranaree University of Technology (SUT), National Science and Technology Development Agency (NSTDA) and National Science, Research and Innovation Fund (NSRF via PMU-B B05F650021), Thailand;
Turkish Energy, Nuclear and Mineral Research Agency (TENMAK), Turkey;
National Academy of  Sciences of Ukraine, Ukraine;
Science and Technology Facilities Council (STFC), United Kingdom;
National Science Foundation of the United States of America (NSF) and United States Department of Energy, Office of Nuclear Physics (DOE NP), United States of America.
In addition, individual groups or members have received support from:
Czech Science Foundation (grant no. 23-07499S), Czech Republic;
FORTE project, reg.\ no.\ CZ.02.01.01/00/22\_008/0004632, Czech Republic, co-funded by the European Union, Czech Republic;
European Research Council (grant no. 950692), European Union;
ICSC - Centro Nazionale di Ricerca in High Performance Computing, Big Data and Quantum Computing, European Union - NextGenerationEU;
Academy of Finland (Center of Excellence in Quark Matter) (grant nos. 346327, 346328), Finland;
Deutsche Forschungs Gemeinschaft (DFG, German Research Foundation) ``Neutrinos and Dark Matter in Astro- and Particle Physics'' (grant no. SFB 1258), Germany.

%% file: Alice_Authorlist_2024-09-12_Opt_C.tex
\begin{flushleft} 
\small

S.~Acharya\,\orcidlink{0000-0002-9213-5329}\,$^{\rm 126}$, 
A.~Agarwal$^{\rm 134}$, 
G.~Aglieri Rinella\,\orcidlink{0000-0002-9611-3696}\,$^{\rm 32}$, 
L.~Aglietta\,\orcidlink{0009-0003-0763-6802}\,$^{\rm 24}$, 
M.~Agnello\,\orcidlink{0000-0002-0760-5075}\,$^{\rm 29}$, 
N.~Agrawal\,\orcidlink{0000-0003-0348-9836}\,$^{\rm 25}$, 
Z.~Ahammed\,\orcidlink{0000-0001-5241-7412}\,$^{\rm 134}$, 
S.~Ahmad\,\orcidlink{0000-0003-0497-5705}\,$^{\rm 15}$, 
S.U.~Ahn\,\orcidlink{0000-0001-8847-489X}\,$^{\rm 71}$, 
I.~Ahuja\,\orcidlink{0000-0002-4417-1392}\,$^{\rm 36}$
% ...

\section*{Affiliation Notes}

$^{\rm I}$ Deceased\\
$^{\rm II}$ Also at: Max-Planck-Institut fur Physik, Munich, Germany\\
$^{\rm III}$ Also at: Italian National Agency for New Technologies, Energy and Sustainable Economic Development (ENEA), Bologna, Italy\\
$^{\rm IV}$ Also at: Dipartimento DET del Politecnico di Torino, Turin, Italy\\
$^{\rm V}$ Also at: Department of Applied Physics, Aligarh Muslim University, Aligarh, India\\
$^{\rm VI}$ Also at: Institute of Theoretical Physics, University of Wroclaw, Poland\\
$^{\rm VII}$ Also at: Facultad de Ciencias, Universidad Nacional Autónoma de México, Mexico City, Mexico\\

\section*{Collaboration Institutes}

$^{1}$ A.I. Alikhanyan National Science Laboratory (Yerevan Physics Institute) Foundation, Yerevan, Armenia\\
$^{2}$ AGH University of Krakow, Cracow, Poland\\
$^{3}$ Bogolyubov Institute for Theoretical Physics, National Academy of Sciences of Ukraine, Kiev, Ukraine\\
$^{4}$ Bose Institute, Department of Physics  and Centre for Astroparticle Physics and Space Science (CAPSS), Kolkata, India\\
$^{5}$ California Polytechnic State University, San Luis Obispo, California, United States\\
$^{6}$ Central China Normal University, Wuhan, China\\
$^{7}$ Centro de Aplicaciones Tecnol\'{o}gicas y Desarrollo Nuclear (CEADEN), Havana, Cuba\\
$^{8}$ Centro de Investigaci\'{o}n y de Estudios Avanzados (CINVESTAV), Mexico City and M\'{e}rida, Mexico\\
$^{9}$ Chicago State University, Chicago, Illinois, United States\\
$^{10}$ China Institute of Atomic Energy, Beijing, China\\
$^{11}$ China University of Geosciences, Wuhan, China\\
$^{12}$ Chungbuk National University, Cheongju, Republic of Korea\\
$^{13}$ Comenius University Bratislava, Faculty of Mathematics, Physics and Informatics, Bratislava, Slovak Republic\\
$^{14}$ Creighton University, Omaha, Nebraska, United States\\
$^{15}$ Department of Physics, Aligarh Muslim University, Aligarh, India\\
$^{16}$ Department of Physics, Pusan National University, Pusan, Republic of Korea\\
$^{17}$ Department of Physics, Sejong University, Seoul, Republic of Korea\\
$^{18}$ Department of Physics, University of California, Berkeley, California, United States\\
$^{19}$ Department of Physics, University of Oslo, Oslo, Norway\\
$^{20}$ Department of Physics and Technology, University of Bergen, Bergen, Norway\\
$^{21}$ Dipartimento di Fisica, Universit\`{a} di Pavia, Pavia, Italy\\
$^{22}$ Dipartimento di Fisica dell'Universit\`{a} and Sezione INFN, Cagliari, Italy\\
$^{23}$ Dipartimento di Fisica dell'Universit\`{a} and Sezione INFN, Trieste, Italy\\
$^{24}$ Dipartimento di Fisica dell'Universit\`{a} and Sezione INFN, Turin, Italy\\
$^{25}$ Dipartimento di Fisica e Astronomia dell'Universit\`{a} and Sezione INFN, Bologna, Italy\\
$^{26}$ Dipartimento di Fisica e Astronomia dell'Universit\`{a} and Sezione INFN, Catania, Italy\\
$^{27}$ Dipartimento di Fisica e Astronomia dell'Universit\`{a} and Sezione INFN, Padova, Italy\\
$^{28}$ Dipartimento di Fisica `E.R.~Caianiello' dell'Universit\`{a} and Gruppo Collegato INFN, Salerno, Italy\\
$^{29}$ Dipartimento DISAT del Politecnico and Sezione INFN, Turin, Italy\\
$^{30}$ Dipartimento di Scienze MIFT, Universit\`{a} di Messina, Messina, Italy\\
$^{31}$ Dipartimento Interateneo di Fisica `M.~Merlin' and Sezione INFN, Bari, Italy\\
$^{32}$ European Organization for Nuclear Research (CERN), Geneva, Switzerland\\
$^{33}$ Faculty of Electrical Engineering, Mechanical Engineering and Naval Architecture, University of Split, Split, Croatia\\
$^{34}$ Faculty of Nuclear Sciences and Physical Engineering, Czech Technical University in Prague, Prague, Czech Republic\\
$^{35}$ Faculty of Physics, Sofia University, Sofia, Bulgaria\\
$^{36}$ Faculty of Science, P.J.~\v{S}af\'{a}rik University, Ko\v{s}ice, Slovak Republic\\
$^{37}$ Faculty of Technology, Environmental and Social Sciences, Bergen, Norway\\
$^{38}$ Frankfurt Institute for Advanced Studies, Johann Wolfgang Goethe-Universit\"{a}t Frankfurt, Frankfurt, Germany\\
$^{39}$ Fudan University, Shanghai, China\\
$^{40}$ Gangneung-Wonju National University, Gangneung, Republic of Korea\\
$^{41}$ Gauhati University, Department of Physics, Guwahati, India\\
$^{42}$ Helmholtz-Institut f\"{u}r Strahlen- und Kernphysik, Rheinische Friedrich-Wilhelms-Universit\"{a}t Bonn, Bonn, Germany\\
$^{43}$ Helsinki Institute of Physics (HIP), Helsinki, Finland\\
$^{44}$ High Energy Physics Group,  Universidad Aut\'{o}noma de Puebla, Puebla, Mexico\\
$^{45}$ Horia Hulubei National Institute of Physics and Nuclear Engineering, Bucharest, Romania\\
$^{46}$ HUN-REN Wigner Research Centre for Physics, Budapest, Hungary\\
$^{47}$ Indian Institute of Technology Bombay (IIT), Mumbai, India\\
$^{48}$ Indian Institute of Technology Indore, Indore, India\\
$^{49}$ INFN, Laboratori Nazionali di Frascati, Frascati, Italy\\
$^{50}$ INFN, Sezione di Bari, Bari, Italy\\
$^{51}$ INFN, Sezione di Bologna, Bologna, Italy\\
$^{52}$ INFN, Sezione di Cagliari, Cagliari, Italy\\
$^{53}$ INFN, Sezione di Catania, Catania, Italy\\
$^{54}$ INFN, Sezione di Padova, Padova, Italy\\
$^{55}$ INFN, Sezione di Pavia, Pavia, Italy\\
$^{56}$ INFN, Sezione di Torino, Turin, Italy\\
$^{57}$ INFN, Sezione di Trieste, Trieste, Italy\\
$^{58}$ Inha University, Incheon, Republic of Korea\\
$^{59}$ Institute for Gravitational and Subatomic Physics (GRASP), Utrecht University/Nikhef, Utrecht, Netherlands\\
$^{60}$ Institute of Experimental Physics, Slovak Academy of Sciences, Ko\v{s}ice, Slovak Republic\\
$^{61}$ Institute of Physics, Homi Bhabha National Institute, Bhubaneswar, India\\
$^{62}$ Institute of Physics of the Czech Academy of Sciences, Prague, Czech Republic\\
$^{63}$ Institute of Space Science (ISS), Bucharest, Romania\\
$^{64}$ Institut f\"{u}r Kernphysik, Johann Wolfgang Goethe-Universit\"{a}t Frankfurt, Frankfurt, Germany\\
$^{65}$ Instituto de Ciencias Nucleares, Universidad Nacional Aut\'{o}noma de M\'{e}xico, Mexico City, Mexico\\
$^{66}$ Instituto de F\'{i}sica, Universidade Federal do Rio Grande do Sul (UFRGS), Porto Alegre, Brazil\\
$^{67}$ Instituto de F\'{\i}sica, Universidad Nacional Aut\'{o}noma de M\'{e}xico, Mexico City, Mexico\\
$^{68}$ iThemba LABS, National Research Foundation, Somerset West, South Africa\\
$^{69}$ Jeonbuk National University, Jeonju, Republic of Korea\\
$^{70}$ Johann-Wolfgang-Goethe Universit\"{a}t Frankfurt Institut f\"{u}r Informatik, Fachbereich Informatik und Mathematik, Frankfurt, Germany\\
$^{71}$ Korea Institute of Science and Technology Information, Daejeon, Republic of Korea\\
$^{72}$ Laboratoire de Physique Subatomique et de Cosmologie, Universit\'{e} Grenoble-Alpes, CNRS-IN2P3, Grenoble, France\\
$^{73}$ Lawrence Berkeley National Laboratory, Berkeley, California, United States\\
$^{74}$ Lund University Department of Physics, Division of Particle Physics, Lund, Sweden\\
$^{75}$ Nagasaki Institute of Applied Science, Nagasaki, Japan\\
$^{76}$ Nara Women{'}s University (NWU), Nara, Japan\\
$^{77}$ National and Kapodistrian University of Athens, School of Science, Department of Physics , Athens, Greece\\
$^{78}$ National Centre for Nuclear Research, Warsaw, Poland\\
$^{79}$ National Institute of Science Education and Research, Homi Bhabha National Institute, Jatni, India\\
$^{80}$ National Nuclear Research Center, Baku, Azerbaijan\\
$^{81}$ National Research and Innovation Agency - BRIN, Jakarta, Indonesia\\
$^{82}$ Niels Bohr Institute, University of Copenhagen, Copenhagen, Denmark\\
$^{83}$ Nikhef, National institute for subatomic physics, Amsterdam, Netherlands\\
$^{84}$ Nuclear Physics Group, STFC Daresbury Laboratory, Daresbury, United Kingdom\\
$^{85}$ Nuclear Physics Institute of the Czech Academy of Sciences, Husinec-\v{R}e\v{z}, Czech Republic\\
$^{86}$ Oak Ridge National Laboratory, Oak Ridge, Tennessee, United States\\
$^{87}$ Ohio State University, Columbus, Ohio, United States\\
$^{88}$ Physics department, Faculty of science, University of Zagreb, Zagreb, Croatia\\
$^{89}$ Physics Department, Panjab University, Chandigarh, India\\
$^{90}$ Physics Department, University of Jammu, Jammu, India\\
$^{91}$ Physics Program and International Institute for Sustainability with Knotted Chiral Meta Matter (WPI-SKCM$^{2}$), Hiroshima University, Hiroshima, Japan\\
$^{92}$ Physikalisches Institut, Eberhard-Karls-Universit\"{a}t T\"{u}bingen, T\"{u}bingen, Germany\\
$^{93}$ Physikalisches Institut, Ruprecht-Karls-Universit\"{a}t Heidelberg, Heidelberg, Germany\\
$^{94}$ Physik Department, Technische Universit\"{a}t M\"{u}nchen, Munich, Germany\\
$^{95}$ Politecnico di Bari and Sezione INFN, Bari, Italy\\
$^{96}$ Research Division and ExtreMe Matter Institute EMMI, GSI Helmholtzzentrum f\"ur Schwerionenforschung GmbH, Darmstadt, Germany\\
$^{97}$ Saga University, Saga, Japan\\
$^{98}$ Saha Institute of Nuclear Physics, Homi Bhabha National Institute, Kolkata, India\\
$^{99}$ School of Physics and Astronomy, University of Birmingham, Birmingham, United Kingdom\\
$^{100}$ Secci\'{o}n F\'{\i}sica, Departamento de Ciencias, Pontificia Universidad Cat\'{o}lica del Per\'{u}, Lima, Peru\\
$^{101}$ Stefan Meyer Institut f\"{u}r Subatomare Physik (SMI), Vienna, Austria\\
$^{102}$ SUBATECH, IMT Atlantique, Nantes Universit\'{e}, CNRS-IN2P3, Nantes, France\\
$^{103}$ Sungkyunkwan University, Suwon City, Republic of Korea\\
$^{104}$ Suranaree University of Technology, Nakhon Ratchasima, Thailand\\
$^{105}$ Technical University of Ko\v{s}ice, Ko\v{s}ice, Slovak Republic\\
$^{106}$ The Henryk Niewodniczanski Institute of Nuclear Physics, Polish Academy of Sciences, Cracow, Poland\\
$^{107}$ The University of Texas at Austin, Austin, Texas, United States\\
$^{108}$ Universidad Aut\'{o}noma de Sinaloa, Culiac\'{a}n, Mexico\\
$^{109}$ Universidade de S\~{a}o Paulo (USP), S\~{a}o Paulo, Brazil\\
$^{110}$ Universidade Estadual de Campinas (UNICAMP), Campinas, Brazil\\
$^{111}$ Universidade Federal do ABC, Santo Andre, Brazil\\
$^{112}$ Universitatea Nationala de Stiinta si Tehnologie Politehnica Bucuresti, Bucharest, Romania\\
$^{113}$ University of Cape Town, Cape Town, South Africa\\
$^{114}$ University of Derby, Derby, United Kingdom\\
$^{115}$ University of Houston, Houston, Texas, United States\\
$^{116}$ University of Jyv\"{a}skyl\"{a}, Jyv\"{a}skyl\"{a}, Finland\\
$^{117}$ University of Kansas, Lawrence, Kansas, United States\\
$^{118}$ University of Liverpool, Liverpool, United Kingdom\\
$^{119}$ University of Science and Technology of China, Hefei, China\\
$^{120}$ University of South-Eastern Norway, Kongsberg, Norway\\
$^{121}$ University of Tennessee, Knoxville, Tennessee, United States\\
$^{122}$ University of the Witwatersrand, Johannesburg, South Africa\\
$^{123}$ University of Tokyo, Tokyo, Japan\\
$^{124}$ University of Tsukuba, Tsukuba, Japan\\
$^{125}$ Universit\"{a}t M\"{u}nster, Institut f\"{u}r Kernphysik, M\"{u}nster, Germany\\
$^{126}$ Universit\'{e} Clermont Auvergne, CNRS/IN2P3, LPC, Clermont-Ferrand, France\\
$^{127}$ Universit\'{e} de Lyon, CNRS/IN2P3, Institut de Physique des 2 Infinis de Lyon, Lyon, France\\
$^{128}$ Universit\'{e} de Strasbourg, CNRS, IPHC UMR 7178, F-67000 Strasbourg, France, Strasbourg, France\\
$^{129}$ Universit\'{e} Paris-Saclay, Centre d'Etudes de Saclay (CEA), IRFU, D\'{e}partment de Physique Nucl\'{e}aire (DPhN), Saclay, France\\
$^{130}$ Universit\'{e}  Paris-Saclay, CNRS/IN2P3, IJCLab, Orsay, France\\
$^{131}$ Universit\`{a} degli Studi di Foggia, Foggia, Italy\\
$^{132}$ Universit\`{a} del Piemonte Orientale, Vercelli, Italy\\
$^{133}$ Universit\`{a} di Brescia, Brescia, Italy\\
$^{134}$ Variable Energy Cyclotron Centre, Homi Bhabha National Institute, Kolkata, India\\
$^{135}$ Warsaw University of Technology, Warsaw, Poland\\
$^{136}$ Wayne State University, Detroit, Michigan, United States\\
$^{137}$ Yale University, New Haven, Connecticut, United States\\
$^{138}$ Yildiz Technical University, Istanbul, Turkey\\
$^{139}$ Yonsei University, Seoul, Republic of Korea\\
$^{140}$ Affiliated with an institute covered by a cooperation agreement with CERN\\
$^{141}$ Affiliated with an international laboratory covered by a cooperation agreement with CERN.\\

\end{flushleft} 